\documentclass[prl,twocolumn,superscriptaddress,showpacs,nofootinbib]{revtex4}

\usepackage{graphicx,amsmath,amssymb,bm}

\newcommand{\beqn}{\begin{equation}}
\newcommand{\eeqn}{\end{equation}}
\newcommand{\bea}{\begin{eqnarray}}
\newcommand{\eea}{\end{eqnarray}}
\newcommand{\ba}{\begin{align}}
\newcommand{\ea}{\end{align}}
\newcommand{\lm}{\Lambda}
\newcommand{\fm}{\, \text{fm}}
\newcommand{\fmi}{\, \text{fm}^{-1}}
\newcommand{\fmiq}{\, \text{fm}^{-3}}
\newcommand{\mev}{\, \text{MeV}}
\newcommand{\vlowk}{V_{{\rm low}\,k}}
\newcommand{\kf}{k_{\rm F}}
\begin{document}

\title{Improved nuclear matter calculations from chiral low-momentum
interactions} 

\author{K.\ Hebeler}
\affiliation{TRIUMF, 4004 Wesbrook Mall, Vancouver, BC, V6T 2A3,
Canada}
\affiliation{Department of Physics, The Ohio State University,
Columbus, OH 43210, USA}
\author{S.\ K.\ Bogner}
\affiliation{National Superconducting Cyclotron Laboratory and
Department of Physics and Astronomy, Michigan State University,
East Lansing, MI 48844, USA}
\author{R.\ J.\ Furnstahl}
\affiliation{Department of Physics, The Ohio State University,
Columbus, OH 43210, USA}
\author{A.\ Nogga}
\affiliation{Institute for Advanced Simulations,
Institut f\"ur Kernphysik and J\"ulich Centre for Hadron Physics,
Forschungszentrum J\"ulich, 52425 J\"ulich, Germany}
\author{A.\ Schwenk}
\affiliation{ExtreMe Matter Institute EMMI, GSI Helmholtzzentrum f\"ur
Schwerionenforschung GmbH, 64291 Darmstadt, Germany}
\affiliation{Institut f\"ur Kernphysik, Technische Universit\"at
Darmstadt, 64289 Darmstadt, Germany}

\begin{abstract}
We present new nuclear matter calculations based on low-momentum
interactions derived from chiral effective field theory
potentials. The current calculations use an improved treatment of
the three-nucleon force (3NF) contribution that includes a corrected
combinatorial factor
beyond Hartree-Fock that was omitted in previous nuclear matter
calculations.
We find realistic saturation
properties using parameters fit only to few-body data, but with larger
uncertainty estimates from cutoff dependence and the 3NF
parametrization than in previous calculations.
\end{abstract}


\maketitle

Major advances in nuclear structure theory over the
last decade have been made by 
expanding the reach of few-body calculations that use
microscopic interactions between nucleons.
This progress has unambiguously established the quantitative
role of three-nucleon forces (3NF) for the ground state
and spectra of light nuclei ($A \leqslant 12$)~\cite{GFMC,NCSMchiral}.
Pioneering extensions to larger nuclei reveal new facets
of the 3NF, such as its role in determining the location of the 
neutron dripline~\cite{Otsuka:2009cs,Hagen:2009mm} and 
in elucidating the doubly-magic
nature of $^{48}$Ca~\cite{Holt:2010yb}. Pushing these successes to still
heavier nuclei, which includes most of the table of nuclides, is a fundamental 
challenge for low-energy nuclear physics.

The historical route to heavy nuclei is through infinite nuclear matter,
a theoretical uniform limit that first turns off the Coulomb interaction, 
which otherwise drives heavier stable nuclei toward an imbalance of 
neutrons over protons and eventually instability.
However, predicting nuclear matter based on microscopic internucleon
forces has proved to be an elusive target. In particular, few-body fits
have not sufficiently constrained 3NF contributions around
saturation density such that nuclear matter calculations are predictive.
Nuclear matter saturation is very delicate, with the binding energy
resulting from cancellations of much larger potential and kinetic
energy contributions. When a quantitative reproduction of empirical 
saturation properties has been obtained, it was imposed by hand
through adjusting short-range three-body forces (see, for example,
Refs.~\cite{Akmal:1998cf,Lejeunenm}).

The lack of progress toward controlled nuclear matter calculations has
long been hindered by the difficulty of the nuclear many-body problem
when conventional nuclear potentials are used. The present
calculations continue an alternative approach to nuclear matter using
soft Hamiltonians derived from interactions fit only to few-body ($A
\leqslant 4$) data.  We find realistic saturation properties within
our theoretical uncertainty bounds without adjustment of parameters.
This progress follows by applying several recent developments:
systematic starting Hamiltonians based on chiral effective field
theory (EFT)~\cite{RMP}, renormalization group (RG)
methods~\cite{Vlowk} to soften the short-range repulsion and
short-range tensor components of the initial chiral
interactions 
so that convergence of many-body calculations is vastly
accelerated~\cite{nucmatt,NCSM,Sonia}, and a new 3NF fitting procedure
to the $^4$He radius rather than the binding energy~\cite{NCSMchiral}.
(Alternative expansions using chiral interactions are described in Refs.~\cite{RMP,Kaiser,Lacour}).
The calculations here also employ an improved treatment
of the 3NF contribution in many-body perturbation theory compared to
Refs.~\cite{nucmatt, previous_NM}, which includes the full treatment
of 3NF double-exchange diagrams and corrected 3NF combinatorial
factors beyond Hartree-Fock. Note that previous calculations of
neutron matter~\cite{chiralnm,nstar} and finite
nuclei~\cite{Otsuka:2009cs,Holt:2010yb} are not affected.

\begin{figure*}[t]
\begin{center}
\includegraphics[scale=0.5,clip=]{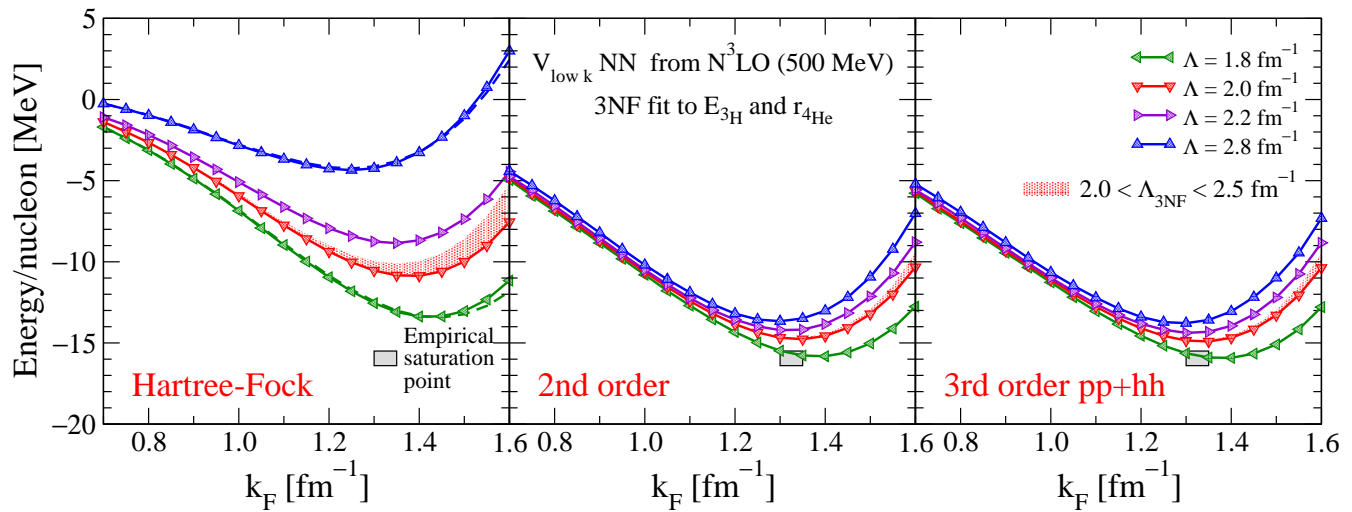}
\end{center}
\vspace*{-5mm}
\caption{(Color online) Nuclear matter energy per particle versus
Fermi momentum $\kf$ at the Hartree-Fock level (left)
and including second-order (middle) and third-order
particle-particle/hole-hole contributions (right),
based on evolved N$^3$LO NN
potentials and 3NF fit to $E_{\rm^3H}$ and $r_{\rm^4He}$. Theoretical
uncertainties are estimated by the NN (lines)/3N (band)
cutoff variations.\label{nm_all}}
\end{figure*}

Our results are summarized in Fig.~\ref{nm_all},
which shows the energy per particle of symmetric matter
as a function of Fermi momentum $\kf$, or the density $\rho
= 2 \kf^3/(3\pi^2)$.
A grey square representing the empirical saturation
point is shown in each of the nuclear matter figures.
Its boundaries reflect the ranges of nuclear matter saturation
properties predicted by phenomenological Skyrme energy functionals that
most accurately reproduce properties of finite nuclei~\cite{Bender}.
Although this determination cannot be completely model independent,
the value is generally accepted for benchmarking infinite matter.
In the future, calculations of the properties of finite nuclei
will allow one to compare directly to experimental data.

The calculations of Fig.~\ref{nm_all} start from the N$^3$LO
nucleon-nucleon (NN) potential (EM $500 \mev$) of
Ref.~\cite{N3LO}. This NN potential is RG-evolved to low-momentum
interactions $\vlowk$ with a smooth $n_{\rm exp}=4$
regulator~\cite{smooth}. For each cutoff $\Lambda$, two couplings that
determine the shorter-range parts of the ${\rm N^2LO}$
3NF~\cite{chiral3N} 
are fit to the $^3$H binding energy and the $^4$He matter radius using
exact Faddeev and Faddeev-Yakubovsky methods as in
Ref.~\cite{Vlowk3N}.  Our 3NF central fit values are given in
Table~\ref{3Nfits}; we estimate that $c_D$ has an uncertainty of approximately $0.4$ due to the uncertainties of the charge radius in $^4$He.
We use a 3NF regulator of the form $\exp[-((p^2 +
3/4 q^2)/\lm_{\rm 3NF}^2)^{n_{\rm exp}}]$ with $n_{\rm
exp} = 4$, where the 3N cutoff $\lm_{\rm 3NF}$ is allowed to vary
independently of the NN cutoff, which probes the sensitivity to 
short-range three-body physics. The shaded regions in
Fig.~\ref{nm_all} show the range of results for 
$2.0 \fmi < \Lambda_{\rm 3NF} < 2.5 \fmi$ at fixed $\Lambda = 2.0 \fmi$.

Nuclear matter is calculated in three approximations: Hartree-Fock
(left), Hartree-Fock plus second-order contributions (middle), and
additionally summing third-order particle-particle and hole-hole
contributions (right).  The technical details regarding the treatment
of the 3NF and the many-body calculation are as for neutron matter in
Ref.~\cite{chiralnm}. We first construct a density-dependent two-body
interaction from the 3NF by summing one particle over occupied states
in the Fermi sea (see also Ref.~\cite{Jeremy}). This conversion
simplifies the many-body calculation significantly and allows the
inclusion of all 3NF double-exchange terms beyond Hartree-Fock, which
were only approximated in Refs.~\cite{previous_NM,nucmatt}.
Furthermore, we have corrected the combinatorial factors 
at the normal-ordered two-body level
of the 3NF from $1/6$ to $1/2$ in diagrams beyond Hartree-Fock
used in these references (see Refs.~\cite{Vlowk,chiralnm} for detailed
discussions of these factors,
which are correctly included in
Refs.~\cite{chiralnm,nstar,Otsuka:2009cs,Holt:2010yb}). To our knowledge,
previous calculations in the literature of nuclear matter using
normal-ordered 3NF contributions need the same correction.

The dashed lines in the left panel 
of Fig.~\ref{nm_all} (for $\Lambda = 1.8 \mev$ and $2.8 \mev$) show the 
exact Hartree-Fock energy in comparison with the results obtained
using the effective two-body interaction (solid lines). 
The excellent agreement supports the use of this 
density-dependent two-body approximation for symmetric nuclear matter.
For the results beyond the Hartree-Fock level we use full
momentum-dependent single-particle Hartree-Fock propagators. 
We have checked that the energies obtained using a self-consistent 
second-order spectrum overlap with the band of curves 
in Fig.~\ref{nm_all}.

The Hartree-Fock results show that nuclear matter is bound even at
this simplest level. A calculation without approximations should be
independent of the cutoffs, so the spread in Fig.~\ref{nm_all} sets
the scale for omitted many-body contributions. The second-order
results show a significant narrowing of this spread over a large
density region. It is encouraging that our results
agree with the empirical saturation point within the uncertainty 
in the many-body calculation and omitted higher-order many-body forces
implied by the cutoff variation (the greater spread compared to
Ref.~\cite{previous_NM} is mostly attributable to the corrected
combinatorial factor). We stress that the cutoff dependence of order
$3 \mev$ around saturation density is small compared to the total size
of the kinetic energy ($\approx 23 \mev$) and potential energy
($\approx -38 \mev$) at this density.
Moreover, the cutoff dependence is smaller at $\kf \approx 1.1 \fmi$,
which resembles more the typical densities in medium-mass to heavy
nuclei ($\rho = 0.11 \fmiq$).
For all cases in the right panel of Fig.~\ref{nm_all}, the
compressibility $K = 175$--$210 \mev$ is in the empirical range.

The inclusion of third-order contributions gives only small changes
from second order except at the lowest densities shown. This is
consistent with nuclear matter being perturbative for low-momentum interactions,
at least in the particle-particle channel~\cite{nucmatt}. The difference at small
densities is not surprising: the presence of a two-body bound state
necessitates a nonperturbative summation in the dilute limit. We note
that below saturation density, the ground state is not a uniform
system, but breaks into clusters (see, for example,
Ref.~\cite{virial}).
 
\begin{table}[b]
\begin{tabular}{c|p{1.225cm}p{1.225cm}|p{1.225cm}p{1.225cm}}
& \multicolumn{2}{c|}{$\vlowk$} & \multicolumn{2}{c}{SRG} \\[0.2mm] \hline
$\Lambda$ or $\lambda/\lm_{\rm 3NF}$ [fm$^{-1}$]
& \multicolumn{1}{c}{$c_D$} & \multicolumn{1}{c|}{$c_E$} 
& \multicolumn{1}{c}{$c_D$} & \multicolumn{1}{c}{$c_E$} \\[0.2mm] \hline
$1.8/2.0$ (EM $c_i$'s) & $+1.621$ & $-0.143$ & $+1.264$ & $-0.120$ \\
$2.0/2.0$ (EM $c_i$'s) & $+1.705$ & $-0.109$ & $+1.271$ & $-0.131$ \\
$2.0/2.5$ (EM $c_i$'s) & $+0.230$ & $-0.538$ & $-0.292$ & $-0.592$ \\
$2.2/2.0$ (EM $c_i$'s) & $+1.575$ & $-0.102$ & $+1.214$ & $-0.137$ \\
$2.8/2.0$ (EM $c_i$'s) & $+1.463$ & $-0.029$ & $+1.278$ & $-0.078$ \\\hline
$2.0/2.0$ (EGM $c_i$'s) & $-4.381$ & $-1.126$ & $-4.828$ & $-1.152$
\\\hline
$2.0/2.0$ (PWA $c_i$'s) & $-2.632$ & $-0.677$ & $-3.007$ & $-0.686$
\end{tabular}
\caption{Results for the $c_D$ and $c_E$ couplings fit to $E_{^3{\rm H}} 
= -8.482 \mev$ and to the point charge radius $r_{^4{\rm He}} = 1.464
\fm$ (based on Ref.~\cite{sick}) for the NN/3N cutoffs and different 
EM/EGM/PWA $c_i$ values used.
For $\vlowk$ (SRG) interactions, the 3NF fits lead to $E_{^4{\rm He}}=
-28.22 \ldots -28.45 \mev$ ($-28.53 \ldots -28.71 \mev$).\label{3Nfits}}
\end{table}

In chiral EFT without explicit Deltas, 3N interactions
start at N$^2$LO~\cite{chiral3N} and their contributions
are given diagrammatically by
\beqn
\parbox[c]{180pt}{
\includegraphics[scale=0.55,clip=]{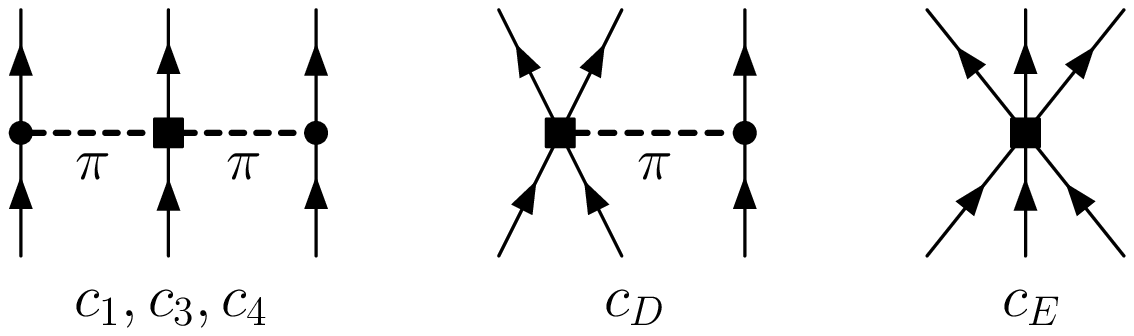}}
\nonumber
\label{3NF}
\eeqn
We assume that the $c_i$ coefficients of the long-range
two-pion-exchange part are not modified by the RG.  At present, we
rely on the N$^2$LO 3NF as a truncated ``basis'' for low-momentum 3N
interactions and determine the $c_D$ and $c_E$ couplings by a fit to
data for all cutoffs~\cite{Vlowk3N}. In the future, fully evolved
three- and four-body forces in momentum space starting from chiral EFT
will be available (see Ref.~\cite{SRG3b} for an application of evolved
3NF in a harmonic-oscillator basis). The fit values of
Table~\ref{3Nfits} are natural and the predicted $^4$He binding
energies are very reasonable. We have also verified that the resulting
3NF becomes perturbative in $A=3,4$ (see also Refs.~\cite{previous_NM,%
nucmatt,Vlowk3N}), i.e., the calculated individual 3NF contributions
are largely unchanged if evaluated for wavefunctions using NN forces
only.

\begin{figure}[t]
\begin{center}
\includegraphics[scale=0.45,clip=]{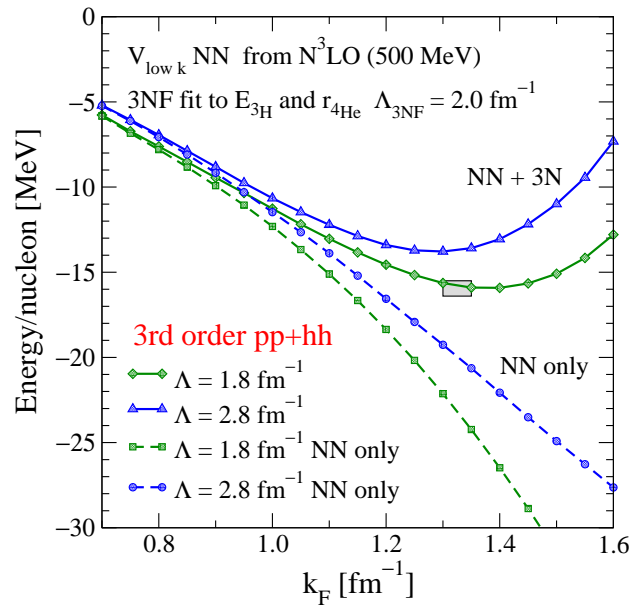}
\end{center}
\vspace{-5mm}
\caption{(Color online) Nuclear matter energy of 
Fig.~\ref{nm_all} at the third-order level compared
to NN-only results for two representative NN cutoffs and a fixed
3N cutoff.\label{NNvs3N}}
\end{figure}

The evolution of the cutoff $\Lambda$ to smaller values is
accompanied by a shift of physics. In particular, effects due
to iterated tensor interactions
are replaced by three-body contributions.
The role of the 3NF for saturation is demonstrated in
Fig.~\ref{NNvs3N}. The two pairs of curves show the difference
between the nuclear matter results for NN-only and NN plus 3N
interactions. It is evident that saturation is driven by the
3NF~\cite{previous_NM,nucmatt}. Even for $\Lambda = 2.8 \fmi$, which is 
similar to the lower cutoffs in chiral EFT 
potentials, saturation is at too high density without
the 3NF. Nevertheless, as in previous results~\cite{previous_NM,nucmatt},
the 3N contributions and the $c_D, c_E$ fits are natural,
and the same is expected for the ratio of three- to four-body
contributions.

\begin{figure}[t]
\begin{center}
\includegraphics[scale=0.45,clip=]{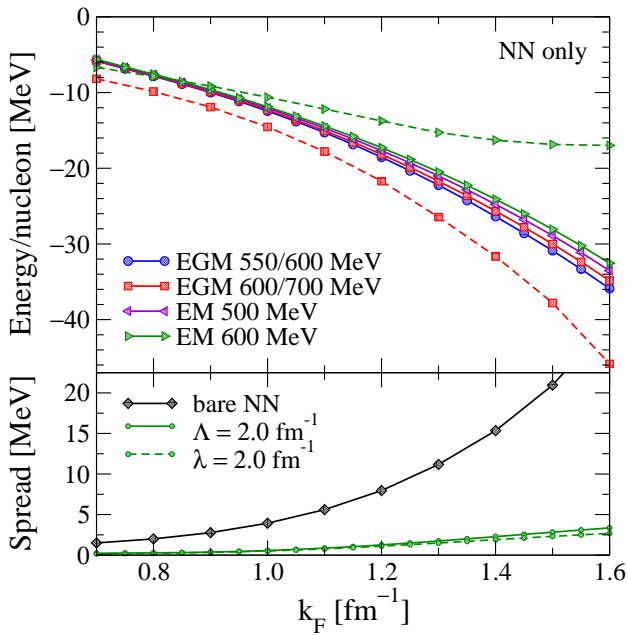}
\end{center}
\vspace*{-5mm}
\caption{(Color online) Nuclear matter NN-only results for 
different chiral N$^3$LO potentials (EM~\cite{N3LO} and 
EGM~\cite{N3LOEGM}). The upper panel shows the third-order results 
for $\vlowk$-evolved interactions at $\Lambda = 2.0\,\rm{fm}^{-1}$
(solid lines) and Brueckner-Hartree-Fock results
for the two unevolved chiral potentials that provide the lowest and
highest energies (dashed lines), EGM 600/700 MeV and EM 600 MeV. 
The lower panel shows the maximal spread of the energy results at 
these two cutoff scales $\Lambda$ for $\vlowk$ and $\lambda$ for 
SRG-evolved NN interactions.} \label{Espread}
\end{figure}

\begin{figure}[t]
\begin{center}
\includegraphics[scale=0.45,clip=]{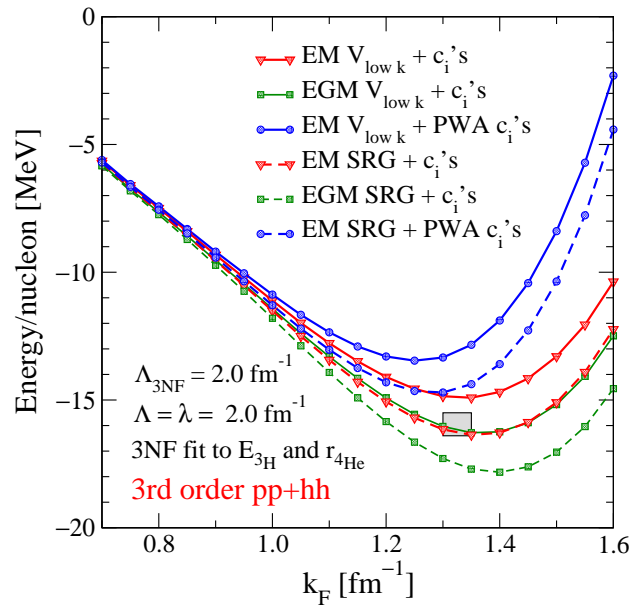}
\end{center}
\vspace*{-5mm}
\caption{(Color online) Nuclear matter energy at the third-order level
comparing low-momentum $\vlowk$ with SRG-evolved chiral
NN interactions for 3NF with different EM/PWA/EGM $c_i$ values
used (see text). \label{nmci}}
\end{figure}

The smooth RG evolution used in the results so far is not the only
choice for low-momentum interactions.  A recent development is the use
of flow equations to evolve Hamiltonians, which we refer to as the
Similarity Renormalization Group (SRG)~\cite{SRG,SRGnuc,Roth:2005pd}.
The flow parameter $\lambda$, which has dimensions of a momentum, is a
measure of the degree of decoupling in momentum space. Few-body
results for roughly the same value of SRG $\lambda$ and smooth
$\vlowk$ $\Lambda$ have been remarkably similar (see, for example,
Ref.~\cite{NCSM}). With either RG method, we can also change the
initial interaction. The results presented so far all start from a
chiral EFT potential at a single order with one choice of EFT
regulator implementation and values. There are several alternatives
available~\cite{RMP,N3LO,N3LOEGM}, which are almost phase-shift
equivalent (but the $\chi^2$ is not equally good up to $E_{\rm lab}
\approx 300 \mev$). Universality for phase-shift equivalent chiral EFT
potentials as $\Lambda$ decreases was shown for smooth-cutoff $\vlowk$
interactions in Ref.~\cite{Vlowk,smooth} in the form of the collapse
of different initial potentials to the same matrix elements in each
partial wave channel. An analogous collapse has been found for N$^3$LO
potentials evolved by the SRG to smaller $\lambda$~\cite{Vlowk}.

Based on this universal collapse for low-momentum interaction matrix
elements it is natural to expect a similar collapse for the energy per
particle in nuclear matter. We consider four different chiral NN
potentials: the N$^3$LO potential by Entem and Machleidt~\cite{N3LO}
for two different cutoffs 500 and 600 MeV, and the N$^3$LO NN
potential by Epelbaum \emph{et al.}~\cite{N3LOEGM} (EGM) for two
different cutoff combinations 550/600 MeV and 600/700 MeV. The results
for the energy are presented in Fig.~\ref{Espread}. The upper panel
shows the energies for $\vlowk$ NN-only interactions derived from
different chiral NN potentials (solid lines) in comparison to
Brueckner-Hartree-Fock (BHF, which means resummed particle-particle
ladder) results based on unevolved chiral potentials (dashed
lines). 
For clarity, we only display the two extreme BHF results. As shown in
the lower panel we find a model dependence of about $13 \mev$ for the
unevolved N$^3$LO potentials around saturation density and
approximately $2 \mev$ for the $\vlowk$ and SRG low-momentum
interactions, comparable to the cutoff variation in
Fig.~\ref{nm_all}. The latter spread reflects the residual 
RG/SRG dependence on the initial interactions.

By supplementing the low-momentum NN interactions with corresponding
3NF's we can probe the sensitivity of the energy to uncertainties in
the $c_i$ coefficients (see also
Refs.~\cite{Bernard:2007zu,chiralnm,nm}). We consider three different
cases: first, we take low-momentum interactions evolved from the
N$^3$LO NN potential EM 500 MeV
(EM $c_i$'s: $c_1 = - 0.81\,{\rm GeV}^{-1}, c_3 = - 3.2\,{\rm
GeV}^{-1}, c_4 = 5.4\,{\rm GeV}^{-1}$), second, low-momentum
interactions from the EGM 550/600 MeV potential (EGM $c_i$'s: $c_1 = -
0.81\,{\rm GeV}^{-1}, c_3 = - 3.4\,{\rm GeV}^{-1}, c_4 = 3.4\,{\rm
GeV}^{-1}$), 
and third, low-momentum interactions from the EM 500 MeV potential
combined with the central $c_i$ values obtained from the NN partial
wave analysis~\cite{PWAci} (PWA $c_i$'s: $c_1 = - 0.76\,{\rm GeV}^{-1},
c_3 = - 4.78\,{\rm GeV}^{-1}, c_4 = 3.96\,{\rm GeV}^{-1}$). The 
fit values for $c_D$ and $c_E$ are given in Table~\ref{3Nfits}.

The resulting nuclear matter energies are shown in
Fig.~\ref{nmci}. For all three cases we find realistic saturation
properties within the theoretical uncertainties implied by the cutoff
dependence shown in Fig.~\ref{nm_all} and the NN
interaction-dependence shown in Fig.~\ref{Espread}. The difference
between $\vlowk$ and SRG results for a given set of $c_i$ is similar
to the NN-only case (see Fig.~\ref{nm_all}), which helps support the
general nature of the 3NF fit. However, the present sensitivity study
can clearly only provide a first estimate for the energy spread due to
uncertainties of the $c_i$ couplings. A more systematic study will
require a correlation analysis based on a larger set of results.

The theoretical errors of our nuclear matter results arise from
truncations in the initial chiral EFT Hamiltonian, the approximation
of the 3NF, and the many-body approximations. Corrections to the
present calculation include higher-order many-body terms, in
particular particle-hole corrections, and contributions from
higher-order many-body forces and from 3NF contributions that cannot
be expressed in terms of density-dependent two-body
interactions. While the improvements in the cutoff dependence suggest
that these corrections are relatively small, an approach such as
coupled cluster theory that can perform a high-level resummation is
ultimately necessary for a robust validation.

While nuclear matter has lost to light nuclei its status as the first
step to nuclear structure, it is still key as a step to heavier nuclei
and astrophysical applications like the structure of neutron
stars~\cite{nstar}. Our results can help with efforts to develop
ab-initio density functional theory (DFT) based on expanding about
nuclear matter~\cite{DME}.  This is analogous to the application of
DFT in quantum chemistry and condensed matter starting with the
uniform electron gas in local-density approximations and adding
constrained derivative corrections. Phenomenological energy
functionals (such as Skyrme) for nuclei have impressive successes but
lack a (quantitative) microscopic foundation based on nuclear forces
and seem to have reached the limits of improvement with the current
form of functionals~\cite{Bertsch:2004us,Kortelainen:2008rp}.  At
present, the theoretical errors of our results, while small on the
scale of the potential energy per particle, are too large to be
quantitatively competitive with existing functionals. 
The implementation of higher-order chiral Hamiltonians and their 
RG evolution can be expected to provide more accurate and reliable predictions.
However, there is also the possibility of fine tuning to
heavy nuclei, of using EFT/RG to guide next-generation functional
forms~\cite{Gebremariam:2010ni,Stoitsov:2010ha}, and of benchmarking
with ab-initio methods for low-momentum interactions. Work in these
directions is in progress.

In summary, we have presented new results for nuclear matter
based on chiral NN and 3N interactions with RG evolution.  The chiral
EFT framework provides a systematic improvable Hamiltonian while the
softening of nuclear forces by RG evolution enhances the convergence
and control of the many-body calculation.  The empirical saturation
point is reproduced within our estimates of uncertainties despite input only
from few-body data.  Because of the fine cancellations, however,
significant reduction of these uncertainties will be needed before
direct DFT calculations of nuclei are competitive.  Nevertheless,
these results are very promising for a unified description of all
nuclei and nuclear matter.

\begin{acknowledgments}
We thank J.\ W.\ Holt for helpful discussions. This work was
supported in part by NSERC, the NSF under Grant Nos.~PHY--0653312,
PHY--0758125 and PHY--1002478, the UNEDF SciDAC Collaboration under DOE Grant
DE-FC02-07ER41457, the Helmholtz Alliance Program of the
Helmholtz Association, contract HA216/EMMI ``Extremes of Density and
Temperature: Cosmic Matter in the Laboratory'' and the DFG through grant SFB 634. 
TRIUMF receives funding via a contribution through the NRC Canada. Part of the
numerical calculations have been performed at the JSC, J\"ulich,
Germany.
\end{acknowledgments}



\begin{thebibliography}{99}

\bibitem{GFMC} S.\ C.\ Pieper, R.\ B.\ Wiringa and J.\ Carlson, Phys.\
Rev.\ C {\bf 70}, 054325 (2004); S.\ C.\ Pieper, Riv.\ Nuovo Cim.\
{\bf 031}, 709 (2008).

\bibitem{NCSMchiral} P.\ Navr\'{a}til, V.\ G.\ Gueorguiev, J.\ P.\ Vary, W.\ E.\ Ormand and A.\ Nogga, Phys.\ Rev.\ Lett.\
{\bf 99}, 042501 (2007); P.\ Navr\'{a}til, S.\ Quaglioni, I.\ Stetcu and B.\ R.\ Barrett, J.\ Phys.\ G {\bf 36} 083101 (2009).

\bibitem{Otsuka:2009cs} T.\ Otsuka, T.\ Suzuki, J.\ D.\ Holt, A.\ Schwenk and Y.\ Akaishi, Phys.\ Rev.\ Lett.\ {\bf 105}, 032501 (2010).

\bibitem{Hagen:2009mm} G.\ Hagen, T.\ Papenbrock, D.\ J.\ Dean, M.\ Hjorth-Jensen and B.\ Velamur Asokan, Phys.\ Rev.\ C {\bf 80}, 021306 (2009).

\bibitem{Holt:2010yb} J.\ D.\ Holt, T.\ Otsuka, A.\ Schwenk and T.\ Suzuki, arXiv:1009.5984.

\bibitem{Akmal:1998cf} A.\ Akmal, V.\ R.\ Pandharipande and D.\ G.\ 
Ravenhall, Phys.\ Rev.\ C {\bf 58}, 1804 (1998).

\bibitem{Lejeunenm} A.\ Lejeune, U.\ Lombardo and W.\ Zuo, Phys.\ Lett.\
B {\bf 477}, 45 (2000).

\bibitem{RMP} E.\ Epelbaum, H.-W.\ Hammer and U.-G.\ Mei{\ss}ner,
\rmp {\bf 81}, 1773 (2009).

\bibitem{Vlowk} S.\ K.\ Bogner, R.\ J.\ Furnstahl and
A.\ Schwenk, Prog.\ Part.\ Nucl.\ Phys. {\bf 65}, 94 (2010).

\bibitem{nucmatt} S.\ K.\ Bogner, A.\ Schwenk, R.\ J.\ Furnstahl and A.\ Nogga, Nucl.\ Phys.\ A {\bf 763}, 59 (2005).

\bibitem{NCSM} S.\ K.\ Bogner, R.\ J.\ Furnstahl, P.\ Maris, R.\ J.\ Perry, A.\ Schwenk and J.\ P.\ Vary, Nucl.\ Phys.\ A {\bf 801}, 21 (2008).

\bibitem{Sonia} S.\ Bacca, A.\ Schwenk, G.\ Hagen and T.\ Papenbrock, Eur.\ Phys.\ J.\ A {\bf 42}, 553 (2009).

\bibitem{Kaiser} N.\ Kaiser, S.\ Fritsch and W.\ Weise, Nucl.\ Phys.\ A {\bf 697}, 255 (2002).

\bibitem{Lacour} A.\ Lacour, J.\ A.\ Oller and U.-G.\ Mei{\ss}ner, Ann.\ Phys.\ {\bf 326}, 241 (2011).

\bibitem{previous_NM} S.\ K.\ Bogner, R.\ J.\ Furnstahl, A.\ Nogga and A.\ Schwenk, arXiv:0903.3366.

\bibitem{chiralnm} K.\ Hebeler and A.\ Schwenk, Phys.\ Rev.\ C 
{\bf 82}, 014314 (2010).

\bibitem{nstar} K.\ Hebeler, J.\ M.\ Lattimer, C.\ J.\ Pethick and A.\ Schwenk, Phys.\ Rev.\ Lett.\ {\bf 105}, 161102 (2010).

\bibitem{Bender} M.\ Bender, P.-H.\ Heenen and P.-G.\ Reinhard, Rev.\ Mod.\ Phys.\ {\bf 75}, 121 (2003).

\bibitem{N3LO} D.\ R.\ Entem and R.\ Machleidt, Phys.\ Rev.\ C {\bf 68},
041001(R) (2003).

\bibitem{smooth} S.\ K.\ Bogner, R.\ J.\ Furnstahl, S.\ Ramanan and A.\ Schwenk, Nucl.\ Phys.\ A {\bf 784}, 79 (2007); K.\ Hebeler, A.\ Schwenk and B.\ Friman,
Phys.\ Lett.\ B {\bf 648}, 176 (2007).

\bibitem{chiral3N} U.\ van Kolck, Phys.\ Rev.\ C {\bf 49}, 2932 (1994);
E.\ Epelbaum, A.\ Nogga, W.\ Gl\"ockle, H.\ Kamada, U.-G. Mei{\ss}ner and H.\ Witala, Phys.\ Rev.\ C {\bf 66}, 064001 (2002).

\bibitem{Vlowk3N} A.\ Nogga, S.\ K.\ Bogner and A.\ Schwenk, Phys.\ Rev.\
C {\bf 70}, 061002(R) (2004).

\bibitem{Jeremy} J.\ W.\ Holt, N.\ Kaiser and W.\ Weise, Phys.\ Rev.\
C {\bf 81}, 024002 (2010).

\bibitem{virial} C.\ J.\ Horowitz and A.\ Schwenk, Nucl.\ Phys.\ A
{\bf 776}, 55 (2006).

\bibitem{SRG3b} S.\ K.\ Bogner, R.\ J.\ Furnstahl and R.\ J.\ Perry,
Ann.\ Phys.\ {\bf 323}, 1478 (2008); E.\ D.\ Jurgenson, P.\ Navratil
and R.\ J.\ Furnstahl, Phys.\ Rev.\ Lett.\ {\bf 103}, 082501 (2009).

\bibitem{sick} I.\ Sick, Phys.\ Rev.\ C {\bf 77}, 041302 (2008).

\bibitem{SRG} S.\ D.\ Glazek and K.\ G.\ Wilson, Phys.\ Rev.\ D {\bf 48},
5863 (1993); F.\ Wegner, Ann.\ Phys.\ (Leipzig) {\bf 3}, 77 (1994).

\bibitem{SRGnuc} S.\ K.\ Bogner, R.\ J.\ Furnstahl and R.\ J.\ Perry,
Phys.\ Rev.\ C {\bf 75}, 061001(R) (2007).

\bibitem{Roth:2005pd} An alternative non-RG use of unitary transformations
to reduce correlations in many-body wave functions is described in
R.\ Roth, H.\ Hergert, P.\ Papakonstantinou, T.\ Neff and H.\ Feldmeier, Phys.\ Rev.\ C {\bf 72}, 034002 (2005), and references therein.

\bibitem{N3LOEGM} E.\ Epelbaum, W.\ Gl\"ockle and U.-G.\ Mei{\ss}ner,
Nucl.\ Phys.\ A {\bf 747}, 362 (2005).

\bibitem{nm} L.\ Tolos, B.\ Friman and A.\ Schwenk, Nucl.\ Phys.\ A
{\bf 806}, 105 (2008).

\bibitem{Bernard:2007zu} V.\ Bernard, Prog.\ Part.\ Nucl.\ Phys.\
{\bf 60}, 82 (2008).

\bibitem{PWAci} M.\ C.\ M.\ Rentmeester, R.\ G.\ E.\ Timmermans and
J.\ J.\ de Swart, Phys.\ Rev.\  C {\bf 67}, 044001 (2003).

\bibitem{DME} S.\ K.\ Bogner, R.\ J.\ Furnstahl and L.\ Platter,
Eur.\ Phys.\ J.\  A {\bf 39}, 219 (2009). B.\ Gebremariam,
T.\ Duguet and S.\ K.\ Bogner, Phys. Rev. C {\bf 82}, 014305 (2010).

\bibitem{Bertsch:2004us} G.\ F.\ Bertsch, B.\ Sabbey and M.\ Uusnakki,
Phys.\ Rev.\  C {\bf 71}, 054311 (2005).

\bibitem{Kortelainen:2008rp} M.\ Kortelainen, J.\ Dobaczewski, K.\ Mizuyama and J.\ Toivanen, Phys.\ Rev.\ C {\bf 77}, 064307 (2008).

\bibitem{Gebremariam:2010ni} B.\ Gebremariam, S.\ K.\ Bogner and
T.\ Duguet, Nucl.\ Phys.\ A {\bf 851}, 17 (2011) 

\bibitem{Stoitsov:2010ha} M.\ Stoitsov, M.\ Kortelainen, S.\ K.\ Bogner, T.\ Duguet, R.\ J.\ Furnstahl, B.\ Gebremariam and N.\ Schunck, Phys.\ Rev.\ C {\bf 82}, 054307 (2010).

\end{thebibliography}
\end{document}